\documentclass[prb,twocolumn,showpacs,preprintnumbers,amsmath,amssymb]{revtex4}
\usepackage{graphicx}% Include figure files
\usepackage{dcolumn}% Align table columns on decimal point
\usepackage{bm}% bold math

\begin{document}

\title{Temperature induced shift in the pinch-off voltage of mesoscopic devices}

\author{
Adam Espe Hansen$^{*,\dagger}$,
Anders Kristensen$^{*,\ddagger}$, and
Henrik Bruus$^{*,\ddagger}$}
\affiliation{
$^{*}$Niels Bohr Institute, University of Copenhagen, Universitetsparken 5, DK-2100 Copenhagen\\
$^{\dagger}$Institute of Physics, University of Basel, Klingelbergstrasse 82, CH-4056 Basel\\
$^{\ddagger}$Mikroelektronik Centret, Technical University of Denmark, DK-2800 Lyngby\\
{\rm (Submitted to Phys.~Rev.~B, 18 September 2002)}
}

%\date{}

\begin{abstract}
Detailed experimental studies of the conductance of mesoscopic
GaAs devices in the few-mode regime
reveal a novel thermal effect: for temperatures up to at least
10~K the measured gate characteristics, i.e.\ conductance $G$ versus
gate voltage $V_g$, exhibit a systematic downward shift in
gate voltage with increasing temperature. The effect is 'universal',
in the sense that it is observed in different modulation doped
GaAs/GaAlAs heterostructures, in different device geometries, and
using different measurement setups. Our observations indicate that
the effect originates in the surrounding 2D electron gas and
not in the mesoscopic devices themselves.
\end{abstract}

\pacs{73.61.-r, 73.23.-b, 73.23.Ad}
\maketitle

\section{Introduction}

\begin{figure*}
\centerline{
\includegraphics[width=2.0\columnwidth]{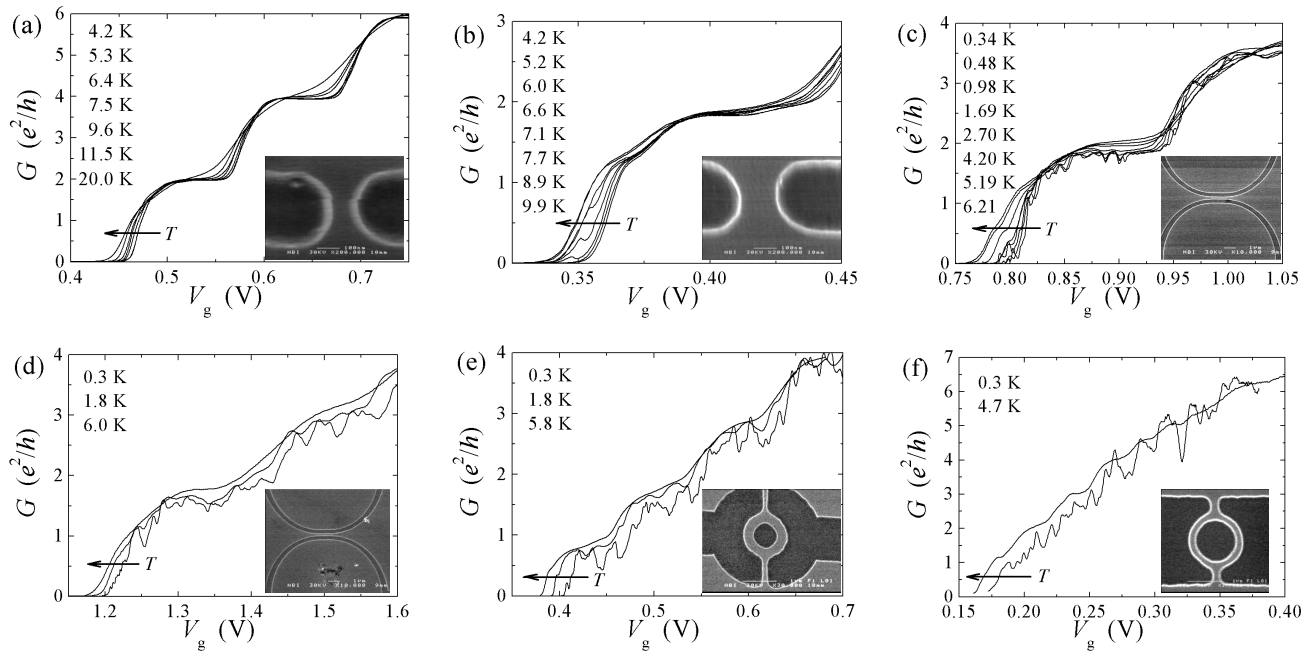}}
\caption{
Six sets of gate characteristics, i.e.\ differential conductance $G$
versus gate voltage $V_g$, from different types of mesoscopic, shallow etched GaAs
devices. In each panel is shown a series of conductance traces taken
at the different tabulated temperatures. The temperature induced
shifts in the pinch-off voltages are indicated by the horizontal
arrows. Scanning electron microscope pictures of the samples are
shown as insets. Panel
(a) is a QPC with a width of 70~nm;
(b) is a QPC with a width of 150~nm;
(c) is a quantum wire with a length of 1000~nm and a width of 150~nm;
(d) is a quantum wire with a length of 1000~nm and a width of 170~nm;
(e) is an AB-ring with an outer (inner) diameter of 1270~nm (670~nm), and
    connecting wires of length 800~nm and width 100~nm;
(f) is an AB-ring with  an outer (inner) diameter of 1500~nm (1300~nm), and
    connecting wires of length 300~nm and width 200~nm.
}
\label{fig1}
\end{figure*}

In numerous studies of ballistic mesoscopic devices reported in the
literature the main effect of increasing the temperature is thermal
smearing \cite{smearing} of
the conductance traces as $k_B T$ becomes comparable with the
characteristic intrinsic energy scale of the device. This is
also the main effect in the six sets of gate characteristics
(conductance $G$ versus gate voltage $V_g$) in
Fig.~\ref{fig1} taken from our own experiments
\cite{Kristensen2000,Hansen2001}
on GaAs quantum point contacts (QPC), quantum wires (QW), and
Aharonov-Bohm (AB) rings.

\begin{figure*}
\includegraphics[width=1.5\columnwidth]{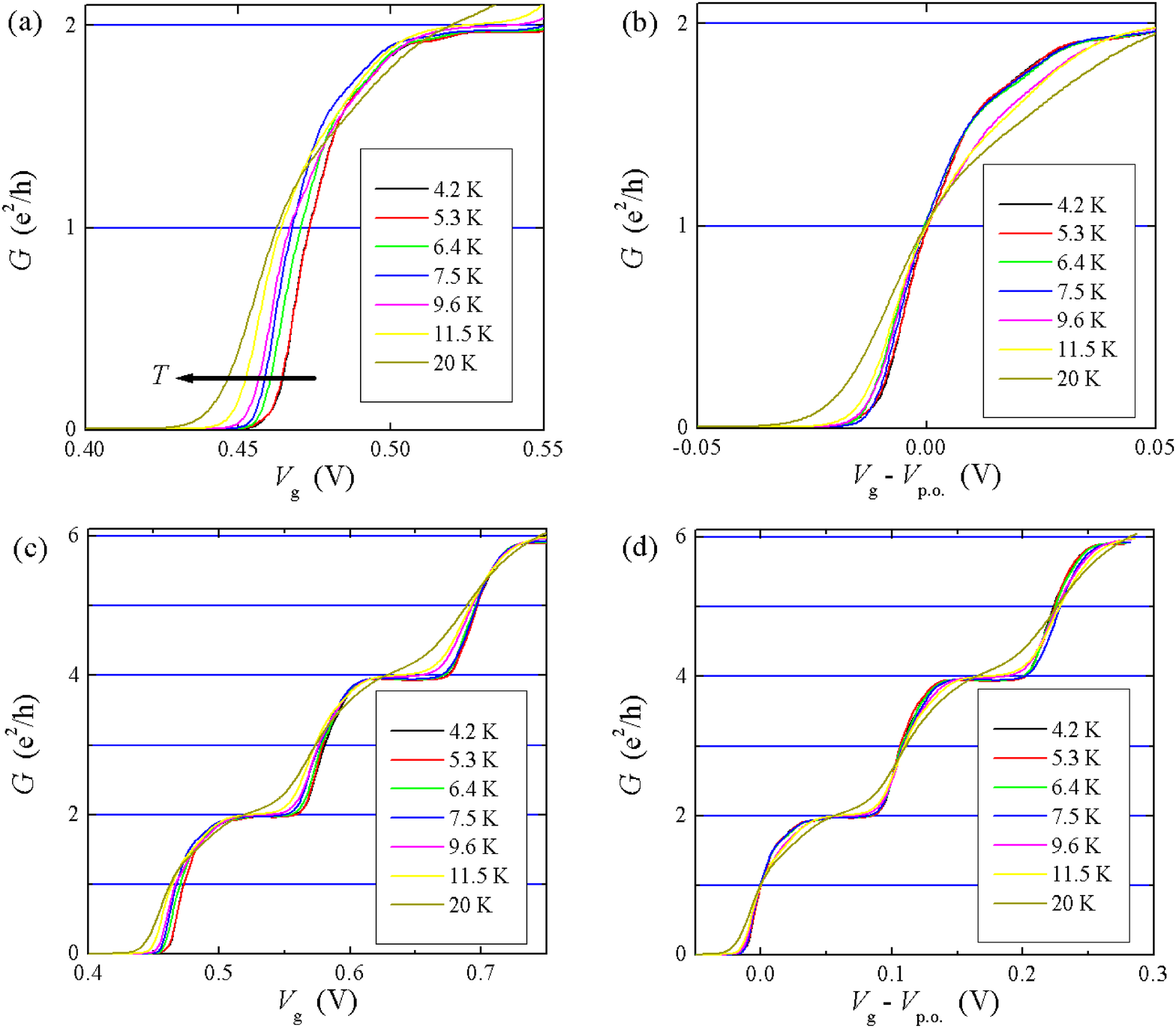}
\caption{The temperature dependence of the conductance traces $G$
  versus $V_g$. (a) A zoom-in on the data from
  Fig.~\ref{fig1}a near pinch-off showing the downward shift of the
  conductance traces with increasing temperature. (b) The data from
  panel (a) shifted to coincide at pinch-off defined as $G=
  e^2/h$. Note that the shifted curves have nearly the same shape
  for $G < e^2/h$. (c) The entire data set showing the first three
  conductance plateaus. (d) The entire data set obtained by the
  shift described in panel (b). Comparing panel (c) and (d) reveals
  that the effect indeed is a shift of the entire trace.
}
\label{fig2}
\end{figure*}

On closer inspection we also observe that the gate characteristics
always exhibit a systematic downward shift in gate voltage with
increasing temperature. We observe this shift in all our
shallow-etched mesoscopic devices independent of the sample geometry and
of the measurement set-up. This observation is the main topic of
this paper, and to our knowledge it has not been reported
in the literature before. The effect may easily be masked by thermal
smearing of the conductance traces, so to study it in detail it is
important to make the intrinsic energy scale as large as possible.
As in our previous work \cite{Kristensen2000}, we achieve this by
relying on a shallow-etch technique in the fabrication of the mesoscopic
devices. This technique yields particularly strong lateral
confinement allowing for well quantized conductances up to 10~K.
The shift of the conductances traces is most clearly seen in
Figs.~\ref{fig1}a and~\ref{fig1}b,
which are QPCs with large subband-spacings, above 10~meV. Hence,
these samples are chosen for the detailed analysis in the following
sections.

\section{Shallow etched mesocopic devices}

Our devices were fabricated on modulation doped GaAs/GaAlAs
heterostructures with a 2-dimensional electron gas (2DEG) buried
90~nm below the surface. Typical 2DEG mobilities and
densities are 60--100~m$^2$V$^{-1}$s$^{-1}$ and 1.5 -- 3$\times$10$^{15}$
m$^{-2}$, respectively. The samples were processed with a $20 \,
\mathrm{\mu m}$ wide and $100 \, \mathrm{\mu m}$ long Hall bar mesa
fitted with ohmic Au/Ge/Ni contacts to the 2DEG. The mesoscopic devices
were defined on top of the mesas by electron beam lithography and
an approximately 30~nm deep wet shallow-etch. The electron densities
are controlled by gate-electrodes in the form of either Cr/Au
top-gate electrodes, deposited on top of the mesoscopic devices,
or by in-plane side-gate electrodes using the 2DEG outside the
shallow etched trenches. For details of the samples and the sample
fabrication, we refer to Ref.~\onlinecite{Kristensen2000}.

Scanning electron microscope (SEM) pictures of some of our shallow
etched mesoscopic devices are shown as insets in Fig.~\ref{fig1}.
The two devices in Figs.~\ref{fig1}a and~\ref{fig1}b are QPCs,
the two devices in Figs.~\ref{fig1}c and~\ref{fig1}d are QWs,
and the last two devices in Figs.~\ref{fig1}e and~\ref{fig1}f are AB
rings. The QWs are controlled by side-gate electrodes, while the QPCs
and the AB rings were fitted with top-gate electrodes after the SEM
pictures were taken. The electrically
insulating, shallow-etched regions appear dark on the pictures. The
mesoscopic devices are normally pinched-off, and
positive gate voltage $V_g$ is applied to open for electron
transport as seen on Fig.~\ref{fig1}. The gate electrode leakage
current was always negligible (less than 100~pA) in the data
presented here.

The QPCs in Figs.~\ref{fig1}a and \ref{fig1}b, henceforth denoted
device A and B, are studied in detail in the following
sections. They are fabricated from the same GaAs/AlGaAs
heterostructure, with a 2DEG having the density
$2\times10^{15}$~m$^{-2}$ and the mobility 80~m$^2$/Vs. The two QPCs
have etched widths of 70~nm and 150~nm, respectively. On both
devices, a $10 \, \mathrm{\mu m}$ wide Cr/Au top-gate electrode
covers the etched constriction and the neighboring 2DEG.  The
devices were characterized by finite source-drain bias
spectroscopy~\cite{Patel91}, and we found an energy spacing
between the first two subbands of 14~meV and 10~meV, respectively.

The data from sample A and B presented here, were recorded by
immersing the sample into a liquid helium container. Temperatures
above $4.2$~K were obtained with the sample placed at different heights
in the helium vapor above the liquid. The differential
conductance $dI/dV_{\rm sd}$ was measured in a four-terminal voltage
controlled setup, using a standard lock-in technique at 117~Hz.
Here $V_{\rm sd}$ is the source-drain voltage bias and $I$ the
corresponding current. The rms amplitude of the applied $V_{\rm
sd}$ was $20 \, \mathrm{\mu V}$, selected to be much smaller than
$k_B T /e$ to ensure linear response. The dc-component of
$V_{\rm sd}$ was zero with a precision of a few $\mu V$.

Our four-terminal measurement setup eliminates most of the series
resistance from the surrounding 2DEG except for a small residual.
To study the thermally induced translation of
the conductance traces it is not necessary to correct for the
residual series resistance. However, following the procedure
described in Ref.~\onlinecite{Hansen2002} it is easy to do so, and
in fact it was done with the data from sample A in the following
analysis.

\begin{figure*}
\centerline{
\includegraphics[width=2.0\columnwidth]{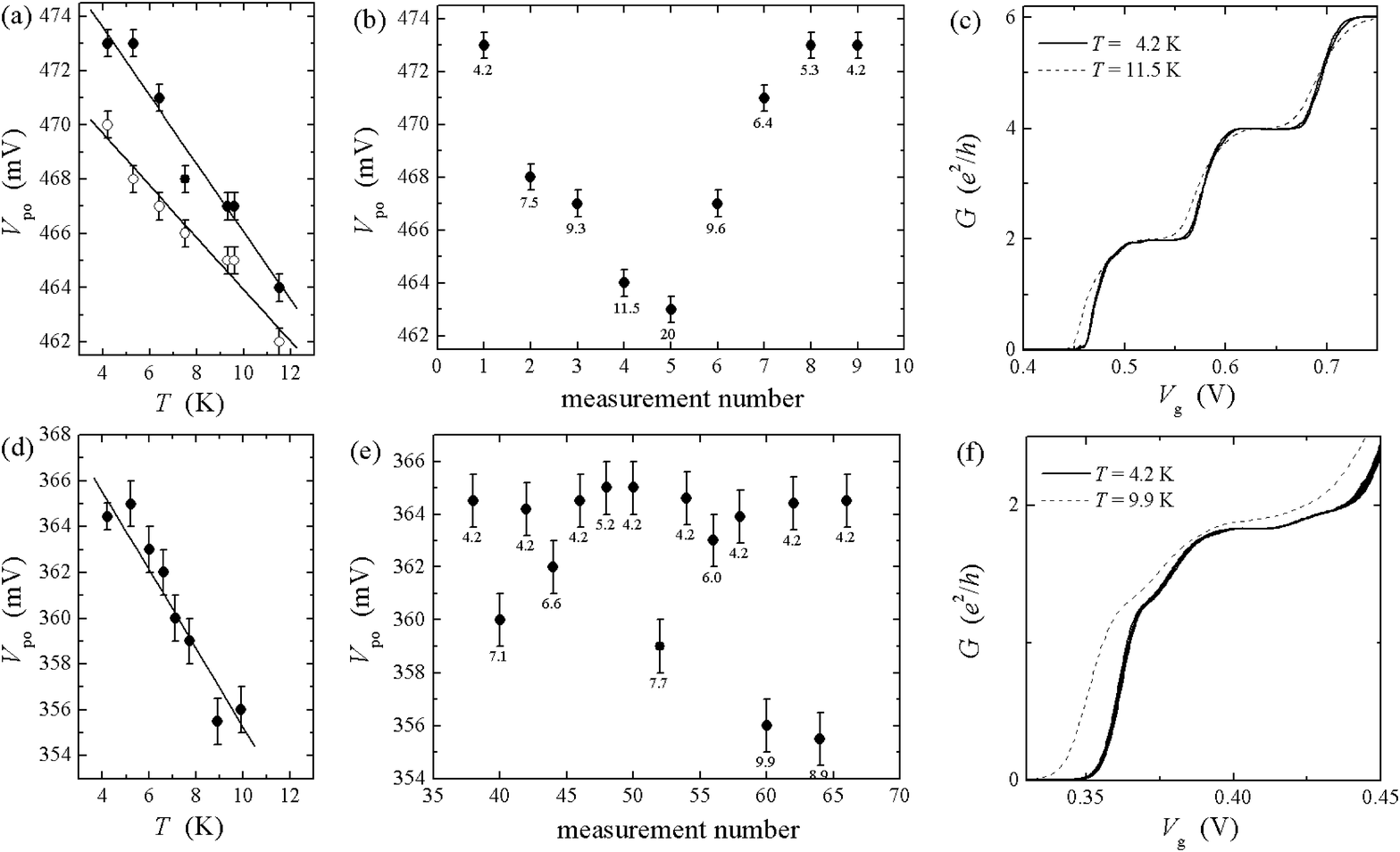}
}
\caption{The temperature dependence of the pinch-off voltages for
  sample A (top row panels) and sample B (bottom row panels). The
  left column, panels (a) and (d), shows graphs of $V_{\rm po}(T)$
  and reveals nearly identical linear behavior, see the text. The
  middle column, panels (b) and (e), shows the measurement sequences
  of the $V_{\rm po}(T)$. The corresponding temperatures $T$ in
  kelvin are written as numbers. Measurements at the base
  temperature 4.2~K are repeated to check temporal stability of the
  devices. In the right column, panel (c) and
  (f), all 4.2~K conductance traces are plotted as black lines, and
  high temperature traces as dotted lines. This illustrates the tiny
  temporal drift of the conductance traces, and compares it to the
  size of the temperature induced shift in pinch-off voltage.
}
\label{fig3}
\end{figure*}

\section{Conductance and pinch-off}

In Fig.~\ref{fig2} we have replotted the conductance data from
Fig.~\ref{fig1}a, i.e.\ from sample A. Fig.~\ref{fig2}a contains
a zoom-in on the first conductance step. It is apparent from this
plot that the conductance traces below $e^2/h$ for the lowest
temperatures have almost identical shapes. The inevitable thermal
smearing is only significant for the highest temperature in the
plot, $T = 20$~K. It seems that the conductance traces are
translated towards lower gate voltages as temperature is
increased. This hypothesis is tested successfully in
Fig.~\ref{fig2}b, where the conductance traces have been shifted
horizontally along the gate voltage axis to coincide at the
conductance value $G = e^2/h$. All the curves (except the $T=20$~K
curve) fall nearly on the same trace for $0 < G \lesssim 1.2 e^2/h$. The
temperature dependence of the conductance is different for
$G \gtrsim 1.2\: e^2/h$ due to the so-called 0.7 structure \cite{Thomas},
an effect we do not study in this work.

We choose the value $G = e^2/h$ as the anchor point in the translation of
the conductance traces for two reasons. Firstly, in the simplest
single-particle models of QPC-conductance this point remains
fixed when temperature is raised until the quantized conductance
is destroyed, i.e., for temperatures smaller than subband spacing
($3.5\: k_B T \lesssim \Delta E$)\cite{Buttiker1990}.
Secondly, it is natural to define the 'pinch-off' voltage
$V_{\rm po}$ at a given temperature as the gate voltage at which the
chemical potential of the reservoirs aligns with the subband edge of
the first 1D subband and the
conductance acquires half its full plateau value, i.e.\
$G(V_{\rm po}) = e^2/h$. The translation we perform is thus
achieved by replacing $V_g$ with $V_g-V_{\rm po}$.

In Figs.~\ref{fig2}c and~\ref{fig2}d we examine the shift in
pinch-off voltage for the higher conductance plateaus. Panel~(c)
and~(d) show the data before and after translation, respectively. On
Fig.~\ref{fig2}d we observe that the translation have turned not
only $G = e^2/h$ into a fix point but also to a good approximation
$G = n e^2/h, n=2,3,4,5$, as predicted by the simple QPC-models
\cite{Buttiker1990}.

To summarize, our data analysis of Fig.~\ref{fig2} reveals that in
addition to smearing, an increased temperature also gives
rise to a systematic downward shift of the conductance traces
towards lower gate voltage. The same conclusion is reached when
analyzing other samples, including the ones shown in
Fig.~\ref{fig1}.

\section{The shift in pinch-off voltage}
\label{po}

To study the pinch-off voltage shift in more detail, we show in
Fig.~\ref{fig3} data from sample A (top row panels) and sample B
(bottom row panels).

In the left column, panels (a) and (d), we have plotted
the measured pinch-off voltage versus temperature as filled
circles. For sample A in panel (a) the open circles show the
'pinch-off' voltage for the second conductance step defined by using
$G = 3e^2/h$ instead of $G = e^2/h$. These data points have been
shifted by $-111$~mV to fit in the figure. We note that all three data sets
show an apparent linear downward shift of the pinch-off voltage with
increasing temperature. The solid lines represent linear fits to the
three data sets, and we find nearly identical slopes $\alpha$,
namely $\alpha^{{}_{}}_{A1} = -(1.3 \pm 0.1)$~mV/K for sample A
first step, $\alpha^{{}_{}}_{A2} = -(1.0 \pm 0.1)$~mV/K for sample A
second step, and $\alpha^{{}_{}}_{B1} = -(1.7 \pm 0.2)$~mV/K for
sample B first step. The nearly identical $\alpha^{{}_{}}_{A1}$ and
$\alpha^{{}_{}}_{A2}$ gives quantitative support for our claim (see
Fig.~\ref{fig2}) that the shift in pinch-off voltage is in fact a shift
of the entire conductance trace. More remarkable is the nearly
identical $\alpha^{{}_{}}_{A1}$ and $\alpha^{{}_{}}_{B1}$, i.e.\ two
differently designed devices from the same heterostructure exhibit
the same shift. This leads to the idea that the temperature induced
shift in the pinch-off voltage is caused by the 2DEG surrounding the
mesoscopic device rather than by the device itself.

The middle and right columns in Fig.~\ref{fig3} are shown to
document our measurement procedure, and to illustrate that
the indeed quite small shift in pinch-off voltage is significant and
reproducible.

The middle column, panel (b) and (e), show the order in which the
data in panel (a) and (d) have been recorded. For each data point
the corresponding temperature in kelvin is indicated. Each sequence
was recorded over several hours to ensure proper thermal equilibrium
at the different temperatures. In order to exclude device
instabilities as an explanation for the observed shifts in pinch-off
voltage, measurements at base temperature 4.2~K were repeated. The
4.2~K data in panels (b) and (e) verify the stability of the
devices.

In the right column, panel (c) and (f), the solid lines are plots
of the whole conductance trace for all base temperature measurements
of panel (b) and (e). These plots show a negligible temporal drift
of the entire conductance trace. The dashed lines in panel (c) and
(f) are conductance traces recorded at high temperature. From the
plots we see that the temperature induced shift of a few mV is
significant compared to the temporal drift.

\section{Suppression of thermal smearing}
\label{smearing}
For conductances in the range $0 < G < e^2/h$ there is a tendency to
suppression of thermal smearing.  While this effect is not central
to the shift in pinch-off voltage, it is very useful in
establishing the very existence of this shift, as demonstrated
clearly in Fig.~\ref{fig2}.

There are several possible explanations for the suppression of
thermal smearing. One is the tunnelling effect through
the QPC barrier. For example, in the quadratic saddle point
approximation the transmission at zero temperature is described by a
Fermi-function with an effective temperature $T_t$. For temperatures
less than $T_t$ the shape of the conductance trace is roughly
independent of temperature and thermal smearing only becomes
significant for $T \gtrsim T_t$.\cite{Buttiker1990}

Another explanation involves Coulomb
interaction between electrons. This interaction gives rise to an
energy scale $E_C = k_B T_C$ that suppresses the thermal smearing for
$T < T_C$. This effect may be seen in the simulations presented in
Ref.~\onlinecite{Hirose}.

\section{Discussion}

We have presented detailed experimental studies of the conductance
of mesoscopic devices in the few-mode regime.
Our data show a novel thermal effect: for temperatures up to at least
10~K the measured conductance traces are systematically shifted
downwards in gate voltage as the temperature is raised. This shift
is 'universal': it is always observed in our shallow-etched mesoscopic devices
independent of sample geometry and measurement set-up. This
apparent 'universality' of the shift leads to the idea
that the effect originates in the physics of the surrounding 2D
electron gas and not in the physics of the mesoscopic device
itself. The observation that different devices from the same wafer
exhibits nearly the same shift lends further support to this idea.

The data presented here are far from sufficient to draw final
conclusions regarding the underlying physics of the temperature
shift. We can rule out non-interacting models. Although these models
are capable of producing a thermally induced change of the anchor
point $G = e^2/h$ by allowing for transmission functions
${\cal T}_n(\varepsilon)$ with non-trivial energy dependencies, such
changes are induced by, and cannot therefore be separated from,
thermal smearing. This is in contrast to the experimental
observations. It may therefore be interesting to consider
interaction effects, e.g., the following possibilities.

The almost pure translation of the conductance traces could most
simply be explained in terms of a temperature dependent shift in the
chemical potential of the surrounding 2DEG. Within the
Landauer-B\"{u}ttiker formalism such a shift measured relative to
the transmission barrier of the mesoscopic device would directly
lead to a translation of the conductance trace, since the
temperature would then play a role similar to a conventional
density-controlling gate electrode. The observed downward shift of
the conductance traces would arise if the chemical potential of the
2DEG rises with increasing temperature. Such a rise could be due to
a gradual reduction of the exchange and correlation energy in the
2DEG.

Another possible explanation could be based on a Coulomb-interaction
induced shift of the subband edge in the mesoscopic device
\cite{Hirose} combined with the existence of a threshold density
controlling the onset of electron transmission. In this case, an
increase in temperature widens the derivative of the
Fermi-distribution, which controls the conductance. The high-energy
tail will then reach the subband edge for a lower chemical
potential. Consequently, the threshold density is reached for lower
chemical potential, i.e.\ for lower gate voltage, as the temperature
is raised.

We would like to emphasize that the 'universality' of the observed
temperature induced shift of the pinch-off voltage means that the
samples shown here are not in any way unique, they simply represent
the samples where the temperature dependence has been checked most
thoroughly by the procedure shown in Fig.~\ref{fig3}.

\section*{Acknowledgements}

We thank C.B. S{\o}rensen for growing the semiconductor
structures. This work was partly supported by the Danish Technical
Research Council (grant no.~9701490) and by the Danish Natural
Science Research Council (grants no.~9502937, 9600548 and
9601677). The III-V materials used in this investigation were made
at the III-V NANOLAB, operated jointly by Research Center COM,
Technical University of Denmark, and the Niels Bohr
Institute, University of Copenhagen.

%\section*{References}


\begin{thebibliography}{9}
\bibitem{smearing} The thermal smearing is spawned by decoherence
  processes and statistical distribution functions.
\end{thebibliography}
\end{document}